\newcommand{\version}{September 14, 2010}
\documentclass[a4paper,twoside,11pt]{article}
\usepackage[bindingoffset=0.5cm,textheight=22.5cm,hdivide={2.8cm,*,2.8cm}, vdivide={*,22cm,*}]{geometry}
\usepackage{amsmath}
\usepackage{amsfonts}
\usepackage{amssymb}
\usepackage{dsfont}
\let\mathbb=\mathds
\usepackage[dvips]{graphicx}
\usepackage[dvips,bookmarksnumbered=true,breaklinks=true]{hyperref}
\usepackage[numbers,square,comma,sort&compress]{natbib}

\newcommand{\uim}{UV/IR mixing}
\newcommand{\nc}{non-com\-mu\-ta\-tive}
\newcommand{\etal}{{\it et al.}}
\newcommand{\eqnref}[1]{Eqn.~(\ref{#1})}		
\newcommand{\figref}[1]{Fig.~\ref{#1}}			
\newcommand{\secref}[1]{Section~\ref{#1}}		
\newcommand{\co}[2]{\left[#1,#2\right]}					
\newcommand{\aco}[2]{\left\{#1,#2\right\}}				
\newcommand{\starco}[2]{\left[ #1\stackrel{\star}{,}#2\right] }		
\newcommand{\var}[2]{\frac{\d #1}{\d #2}}				
\newcommand{\pa}{\partial}						
\newcommand{\ri}{{\rm i}}						
\renewcommand{\k}{\tilde{k}}						
\newcommand{\p}{\tilde{p}}						
\newcommand{\q}{\tilde{q}}						
\newcommand{\Dt}{\widetilde{D}}						
\newcommand{\bc}{\bar{c}}						
\newcommand{\Act}{S}
\renewcommand{\a}{\alpha}
\renewcommand{\b}{\beta}
\newcommand{\g}{\gamma}
\renewcommand{\d}{\delta}
\newcommand{\e}{\epsilon}

\renewcommand{\th}{\theta}
\newcommand{\mth}{\theta} 
\newcommand{\sth}{\varepsilon} 
\newcommand{\st}{\bar{\sigma}}

\newcommand{\m}{\mu}
\newcommand{\n}{\nu}

\renewcommand{\r}{\rho}
\newcommand{\s}{\sigma}
\renewcommand{\t}{\tau}

\newcommand{\G}{\Gamma}

\renewcommand{\L}{\Lambda}

\newcommand{\W}{\Omega}

\newcommand{\inv}[1]{\frac{1}{#1}}				
\newcommand{\tinv}[1]{\tfrac{1}{#1}}
\newcommand{\intk}{\int\limits_{-\infty}^{+\infty}\!\!\frac{\mathrm{d}^4k}{\left(2\pi\right)^4}}
\newcommand{\intx}{\int\!\! {\rm d}^4x}						
\newcommand{\wsq}{\widetilde{\square}}
\newcommand{\ig}{{\rm i}g}
\renewcommand{\intk}{\int\!\! d^4k} 

\newcommand{\bpsi}{\bar{\psi}}
\newcommand{\bB}{\bar{B}}

\newcommand{\ST}{\mathcal{B}_{S}}
\newcommand{\go}{\mathcal{G}}
\newcommand{\ag}{\bar{\mathcal{G}}}
\newcommand{\U}{\mathcal{U}}
\newcommand{\Q}{\mathcal{Q}}

\title{\begin{flushright}
       \small{TUW-09-17}\\\small{UWThPh-2009-14}
       \end{flushright}
\vspace{2em}
On Non-Commutative $U_\star(1)$ Gauge Models and Renormalizability}
\author{Daniel N. Blaschke\footnotemark[1]~\footnotemark[2]~,
Arnold Rofner\footnotemark[1]~, Ren\'e I.P. Sedmik\footnotemark[1]~ \\and Michael Wohlgenannt\footnotemark[1]~}

\date{\version}
\graphicspath{{./figures/}}
\begin{document}
\maketitle
\thispagestyle{empty}
\begin{center}
\renewcommand{\thefootnote}{\fnsymbol{footnote}}
\vspace{-0.3cm}\footnotemark[1]Institute for Theoretical Physics,
Vienna University of Technology\\
Wiedner Hauptstrasse 8-10, A-1040 Vienna (Austria)\\[0.3cm]
\footnotemark[2]Faculty of Physics,
University of Vienna\\
Boltzmanngasse 5, A-1090 Vienna (Austria)\\[0.3cm]
\ttfamily{E-mail: blaschke@hep.itp.tuwien.ac.at, arofner@hep.itp.tuwien.ac.at, sedmik@hep.itp.tuwien.ac.at, miw@hep.itp.tuwien.ac.at}
\vspace{0.5cm}
\end{center}%
\begin{abstract}
Based on our recent findings regarding (non-)renormalizability of {\nc} $U_\star(1)$ gauge theories~\cite{Blaschke:2009c,Blaschke:2009d} we present the construction of a new type of model. By introducing a soft breaking term in such a way that only the bilinear part of the action is modified, no interaction between the gauge sector and auxiliary fields occurs. Demanding in addition that the latter form BRST doublet structures, this leads to a minimally altered {\nc} $U_\star(1)$ gauge model featuring an IR damping behavior. Moreover, the new breaking term is shown to provide the necessary structure in order to absorb the inevitable quadratic IR divergences appearing at one-loop level in theories of this kind. In the present paper we compute Feynman rules, symmetries and results for the vacuum polarization together with the one-loop renormalization of the gauge boson propagator and the three-point functions.
\end{abstract}

\newpage
\tableofcontents

%
\section{Introduction}
When considering quantum field theories on non-commutative spaces, e.g. by employing the so-called Groenewold-Moyal star product, one inevitably has to deal with the infamous UV/IR mixing problem (see \cite{Tanasa:2008d,Rivasseau:2007a,Douglas:2001} for reviews of the topic): New kinds of non-local IR divergences prevent the model from being renormalizable. In fact, so far only some modified scalar field theories on Euclidean non-commutative spaces have been found to be renormalizable by adding new types of terms in the action. The first successful approach was introduced by Grosse and Wulkenhaar~\cite{Grosse:2003}, and proofs of renormalizability have been achieved mainly by utilizing Multiscale Analysis~\cite{Rivasseau:2005a,Rivasseau:2005b}, or formally in the matrix base~\cite{Grosse:2004b}. Recently, and quite independent of former developments, Gurau \etal~\cite{Gurau:2009} introduced a term of the type $\phi\star\inv{\square}\phi$ into the Lagrangian which modifies the theory in the infrared region and, in this way, renders it renormalizable. This was in fact proven to all orders by the authors using Multiscale Analysis. A thorough study of the divergence structure of this model (referred to as $\inv{p^2}$-model), including explicit renormalization at one-loop level~\cite{Blaschke:2008b} has been carried out as well as a computation of the beta functions~\cite{Tanasa:2008a}.

The task of constructing a renormalizable {\nc} gauge theory is even more involved, and although there have been several interesting ideas using additional constraints~\cite{Slavnov:2003}, or generalizing existing scalar models~\cite{Blaschke:2007b,Grosse:2007,Wulkenhaar:2007,Blaschke:2008a}, proofs of renormalizability are still missing. Motivated by the inherent translation invariance and simplicity of the scalar $\inv{p^2}$-model, a possible generalization to gauge theories was introduced in~\cite{Blaschke:2008a} and further discussed in~\cite{Blaschke:2009a,Vilar:2009,Blaschke:2009b}. The main idea was to include a gauge invariant non-local term in the action
\begin{align}
\Act_{\text{nloc}}=\intx\left(\inv{4}F_{\m\n}\star F_{\m\n}+ F_{\m\n}\star\frac{a^2}{\sth^2D^2\Dt^2}\star F_{\m\n}\right)\,,
\label{eq:action_nloc}
\end{align}
where the field strength tensor $F_{\m\n}$ associated to the gauge field $A_\m$ and the covariant derivative $D_\m$ are defined by
\begin{align}
&F_{\m\n}=\partial_\m A_\n -\partial_\n A_\m -\ig \starco{A_{\m}}{A_{\n}}, \nonumber\\
&D_{\m}\cdot = \partial_\m\cdot -\ig \starco{A_\m}{\cdot}, \qquad \text{and}\quad \sth\Dt_{\m}=\sth\mth_{\m\n}D_{\n}\,.
\end{align}
This model is formulated on Euclidean $\mathbb{R}_\mth^4$ with the Moyal-deformed product 
\begin{align}
\starco{x_\m}{x_\n} \equiv  x_{\mu} \star
x_{\nu} -x_{\nu} \star x_{\mu} = \ri \sth\mth_{\mu \nu}\,,
\end{align}
of regular commuting coordinates $x_\mu$. The real parameter $\sth$ has mass dimension -2, rendering the constant antisymmetric matrix $\mth_{\m\n}$ dimensionless. In \eqnref{eq:action_nloc}, the non-local second term implements an IR damping mechanism similar to the one of the corresponding scalar model, but involves an infinite number of vertices in the action~\cite{Blaschke:2008a}. Therefore, two alternatives of localizing that term by introducing additional fields were presented in~\cite{Vilar:2009,Blaschke:2009b}. Initially the intent was to apply the technique of algebraic renormalization, however it turned out that this scheme may not be applicable to {\nc} spaces, as has been discussed recently in Ref.~\cite{Blaschke:2009d}. The reason for these obstacles is that arbitrary powers of the gauge invariant dimensionless operators $\sth^2D^2\Dt^2$ and $\sth\widetilde{F}=\sth\mth_{\m\n}F_{\m\n}$ may appear as counter terms in the effective $n$-loop action. In addition, these models suffer from a high degree of complexity due to a large number of (ghosts and) fields, being introduced in order to implement the demanded symmetries. Finally, due to potentially divergent contributions by auxiliary fields, it is not clear if the models are at all renormalizable.
\vspace{1em}

The aim of the current work is to put forward an alternative action which implements the same\footnote{i.e. the same damping mechanism as in the scalar $\inv{p^2}$-model.} damping mechanism in the gauge field propagator in such a way, that the problems of the former models do not appear. In \secref{sec:def_newmodel}, we present the action of our new model and its (symmetry-) properties. We then discuss one-loop calculations including the respective renormalized propagator in \secref{sec:one-loop} before explaining why we expect no IR problems in higher loop graphs in \secref{sec:higher-loops}.
\vspace{1em}

In the following we will use the abbreviations $\tilde{v}_\m \equiv \mth_{\m\n}v_\n$ for vectors $v$ and $\tilde{M} \equiv \mth_{\m\n}M_{\m\n}$ for matrices $M$. For the deformation, we furthermore consider the simplest block-diagonal form
\begin{align}\label{eq:def-eth}
( \mth_{\mu\nu} )
=\left(\begin{array}{cccc}
0&1&0&0\\
-1&0&0&0\\
0&0&0&1\\
0&0&-1&0
\end{array}\right)\, ,
\end{align}
for the dimensionless matrix describing non-commutativity.

%
\section{Construction of the Action}\label{sec:def_newmodel}
%

Bearing in mind the problems of the previous approaches \cite{Vilar:2009,Blaschke:2009b} to the generalization of the scalar $\inv{p^2}$ model to gauge theories, we now attempt to make a different ansatz. As a starting point we have to take into account that our action should meet the following requirements:
\begin{itemize}
 \item[-] The tree level action should provide a counter term for the quadratic IR divergence in the external momentum $k$
\begin{align}\label{eq:generic-IR-div}
\Pi^{\text{IR}}_{\m\n}(k)&\propto\frac{\k_\m\k_\n}{(\sth\k^2)^2}\,,
\end{align}
which appears at one-loop level in all versions of {\nc} gauge theories on $\th$-deformed spaces.
 \item[-] All relevant propagators should be infrared finite and feature damping factors similar to the scalar $1/p^2$ model~\cite{Gurau:2009}. In this way, IR damping at higher loop orders is implemented without breaking translation invariance.
 \item[-] Any auxiliary fields and related additional ghosts should be decoupled from the gauge sector, i.e. no according interactions should appear so that the physical content is not altered compared to traditional implementations of NC Yang-Mills theory --- see e.g.~\cite{Hayakawa:1999b,Ruiz:2000,Armoni:2000xr}.
 \item[-] The model should be as simple as possible.
\end{itemize}

The present idea is to discard the original approach of adding the non-local term of \eqnref{eq:action_nloc}, and instead to implement the damping entirely within a so-called `soft-breaking' term --- a method well known from the Gribov-Zwanziger action~\cite{Gribov:1978,Zwanziger:1989,Zwanziger:1993,Dudal:2008,Baulieu:2009} in QCD\footnote{In QCD the soft-breaking is introduced in order to restrict the gauge fields to the first Gribov horizon which removes any residual gauge ambiguities, and thereby cures the Gribov problem. In other words, one introduces an additional gauge fixing in the infrared without modifying the ultraviolet region. For details we refer to the extensive literature~\cite{Gribov:1978,Zwanziger:1989,Zwanziger:1993,Dudal:2008,Baulieu:2009}. In the present case we are dealing with a similar problem: the infrared region of our model requires a modification due to {\uim} while the symmetries, which effectively contribute to the renormalizability in the UV, shall not be altered. Hence, we follow here the same strategy as Gribov and Zwanziger.}.

Let us discuss the most important steps leading to our new model. The starting point is the action proposed in Ref.~\cite{Blaschke:2009b}, which has been introduced in order to localize the operator $(D^2\Dt^2)^{-1}$ appearing in \eqnref{eq:action_nloc}. That action consists of the usual Yang-Mills term $\inv{4}\intx F_{\m\n}F_{\m\n}$ plus the terms of \eqref{eq:constr_step0} below. The one proportional to $a$ represents the Gribov-Zwanziger soft breaking, and the second term implements the coupling between auxiliary fields and the gauge sector. Our new proposal is derived in two steps:
\begin{subequations}
\label{eq:brss_constr_evol}
\begin{gather}
\intx \left[\frac{a}{2}\left(B_{\m\n}+\bB_{\m\n}\right)F_{\m\n}-\bB_{\m\n}\,\sth^2\Dt^2D^2B_{\m\n}\right]\label{eq:constr_step0}\,,\\
\phantom{\text{step 1}}\downarrow\text{step 1}\notag\\
\intx \left[\frac{\gamma^3}{2}\left(B_{\m\n}+\bB_{\m\n}\right)\inv{\wsq}F_{\m\n}-\bB_{\m\n}D^2 B_{\m\n}\right]\label{eq:constr_step1}\,,\\
\phantom{\text{step 2}}\downarrow\text{step 2}\notag\\
\intx \left[\frac{\gamma^2}{2}\left(B_{\m\n}+\bB_{\m\n}\right)\inv{\wsq}\left(f_{\m\n}+\s\frac{\mth_{\m\n}}{2}\tilde{f}\right)-\bB_{\m\n}B_{\m\n}\right]\,,\label{eq:constr_step2}\,
\end{gather}
\end{subequations}
with several new definitions to be explained subsequently.

To understand the first step we note that the divergences in the $G^{AB}$, $G^{A\bB}$, $G^{\bB B}$, and $G^{BB}$ propagators (see Ref.~\cite{Blaschke:2009d}) are mainly caused by the appearance of the operator $D^2\Dt^2$ sandwiched between $\bB_{\m\n}$ and $B_{\m\n}$. On the other hand this term is crucial to the construction of the correct damping factor for the gauge boson propagator $G^{AA}$. A detailed analysis~\cite{Sedmik:2009} of the interplay between terms in the action and the resulting propagators leads to the insight that it is possible to move one part ($\Dt^2$) of the problematic operator into the soft breaking term. Aiming to eventually avoid couplings between auxiliary fields and the gauge sector, $\Dt^2$ is furthermore replaced by\footnote{Of course, $\wsq$ is not a gauge invariant operator, but as we will discuss subsequently, BRST invariance of the action will be guaranteed by additional sources.} $\wsq$. Thereby, the desired damping is maintained while the IR divergences in the propagators of the auxiliary fields are eliminated. 
Note also that the dimensionful $\sth$ has been absorbed into the new parameters $\g$ and $\s$, which have mass dimensions $d_m(\g)=1$ and $d_m(\s)=0$.

In step 2, we note that the regularizing effects are solely implemented in the bi-linear part\footnote{Once more, BRST invariance will be discussed subsequently.} of the action, therefore opening the option to reduce the field strength tensor $F_{\m\n}$ in the soft breaking term to its bi-linear part 
\begin{align}\label{eq:def-f}
f_{\m\n}\equiv\partial_\m A_\n-\partial_\n A_\m\,, \qquad 
\tilde{f}\equiv \mth_{\m\n} f_{\m\n}= 2 \tilde{\partial}\cdot A\,.
\end{align}
 Noting furthermore, that the $D^2$ operator in the $\bB/B$ sector is not required any more for the implementation of the damping mechanism, we may entirely omit these covariant derivatives. Consequently, the mass dimensions $d_m$ of the fields $B_{\m\n}$ and $\bB_{\m\n}$ change from 1 to 2. Any interaction (represented by $n$-point functions with $n\geq3$) of $A_\m$ with auxiliary fields and related ghosts is now eliminated. This represents the advantage that no additional Feynman diagrams appear since the auxiliary fields only enter the bi-linear part of the action. Finally, in order to implement a suitable term to absorb the $\theta$-contracted one-loop divergence (see \eqnref{eq:generic-IR-div} and \cite{Blaschke:2009d}), we further modify the soft breaking part by the insertion of the term $\frac{\gamma^2}{4}\s\left(B_{\m\n}+\bB_{\m\n}\right)\inv{\wsq}\mth_{\m\n}\tilde{f}$, resulting in \eqref{eq:constr_step2}. The new parameters $\s$ and $\g$ are intended to receive corrections in the renormalization process, as will be discussed in \secref{sec:renorm}. Obviously the motivation for the new breaking term is of purely technical nature since it simply allows to achieve the requirements we have imposed to the new model at the beginning of this section. Physically, as has been discussed extensively in Refs.~\cite{Vilar:2009,Blaschke:2009b,Sedmik:2009,Rofner:2009}, a soft breaking term in the action (which should be understood as an additional gauge fixing as in the Gribov-Zwanziger case) enables us to modify the behavior at low energy (i.e. to damp IR divergences) while not spoiling the symmetries of the theory in the UV. 
This should be understood as an additional gauge fixing as in the Gribov-Zwanziger case. However, the freedom in the choice of this soft breaking is not yet entirely understood. 
Clearly, due to the {\uim} problem, a non-commutative gauge field model only has the chance to be renormalizable with this or a similar additional IR modification. This is inherently different from the well-known Yang-Mills theory on commutative space and of course requires further investigation in the future. 

In view of these considerations, the following BRST invariant action formulated in Euclidean $\mathbb{R}_\mth^4$ is put forward:
{\allowdisplaybreaks
\begin{align}\label{eq:renormalizable_action}
\Act&=\Act_{\text{inv}}+\Act_{\text{gf}}+\Act_{\text{aux}}+\Act_{\text{soft}}+\Act_{\text{ext}}\,,\nonumber\\
\Act_{\text{inv}}&=\intx\tinv{4}F_{\m\n}F_{\m\n}\,,\nonumber\\
\Act_{\text{gf}}&=\intx\,s\left(\bc\,\pa_\m A_\m\right)=\intx\left(b\,\pa_\m A_\m-\bc\,\pa_\m D_\m c\right)\,,\nonumber\\
\Act_{\text{aux}}&=-\intx\,s\left(\bpsi_{\m\n}B_{\m\n}\right)=\intx\left(-\bB_{\m\n}B_{\m\n}+\bpsi_{\m\n}\psi_{\m\n}\right)\,,\nonumber\\
\Act_{\text{soft}}&=\intx\,s\left[\!\left(\bar{Q}_{\m\n\a\b}B_{\m\n}+Q_{\m\n\a\b}\bar{B}_{\m\n}\right)\inv{\wsq}\left(f_{\a\b}+\s\frac{\mth_{\a\b}}{2}\tilde{f}\right)\right]=\nonumber\\*
&=\intx\bigg[\!\!\left(\bar{J}_{\m\n\a\b}B_{\m\n}+J_{\m\n\a\b}\bar{B}_{\m\n}\right)\!\inv{\wsq}\!\left(f_{\a\b}+\s\frac{\mth_{\a\b}}{2}\tilde{f}\right)\! - \bar{Q}_{\m\n\a\b}\psi_{\m\n}\inv{\wsq}\!\left(f_{\a\b}+\s\frac{\mth_{\a\b}}{2}\tilde{f}\right)\nonumber\\*
& \qquad \qquad -\left(\bar{Q}_{\m\n\a\b}B_{\m\n}+Q_{\m\n\a\b}\bar{B}_{\m\n}\right)\inv{\wsq}\mathop{s}\left(f_{\a\b}+\s\frac{\mth_{\a\b}}{2}\tilde{f}\right)\bigg]\,, \nonumber\\
\Act_{\text{ext}}&=\intx\left(\W^A_\m sA_\m+\W^c sc\right)\,,
\end{align}
where $\wsq = \tilde\partial_\mu \tilde\partial_\mu$, and all products are implicitly assumed to be deformed (i.e. star products). This will also apply to the rest of our paper.
The {\nc} generalization of a $U(1)$ gauge field is denoted by $A_\m$, $\bc$ and $c$ are the (anti-)ghosts and the multiplier field $b$ implements the Landau gauge fixing $\pa_\m A_\m=0$. 
$\W^A_\m$ and $\W^c$ are external sources introduced for the non-linear BRST transformations $sA_\m$ and $sc$, which will be defined shortly. Furthermore, the complex field $B_{\m\n}$ and its conjugate $\bB_{\m\n}$ as well as associated ghosts $\psi_{\m\n}$ and $\bpsi_{\m\n}$ have been introduced into the bilinear part of the action in order to implement the IR damping and BRST invariance (compatible to the `soft breaking' technique). These new unphysical fields do not interact with the gauge field $A_\m$.
Notice, that the Landau gauge fixing can surely be replaced by any other choice, however the freedom in the choice of the soft breaking is not yet entirely understood. 
}

The additional sources $\bar{Q},Q,\bar{J},J$ are needed in order to ensure BRST invariance of $\Act_{\text{soft}}$ in the ultraviolet. At the opposite end of the spectrum (in the infrared) these sources take their `physical values'
\begin{align}\label{JQ-phys_new}
\bar{Q}_{\m\n\a\b}\big|_{\text{phys}}=0\,,\qquad \bar{J}_{\m\n\a\b}\big|_{\text{phys}}=\frac{\g^2}{4}\left(\d_{\m\a}\d_{\n\b}-\d_{\m\b}\d_{\n\a}\right)\,,\nonumber\\
Q_{\m\n\a\b}\big|_{\text{phys}}=0\,,\qquad J_{\m\n\a\b}\big|_{\text{phys}}=\frac{\g^2}{4}\left(\d_{\m\a}\d_{\n\b}-\d_{\m\b}\d_{\n\a}\right)\,,
\end{align} 
see Refs.~\cite{Zwanziger:1993,Dudal:2008,Vilar:2009} for details on this technique. 
The action \eqref{eq:renormalizable_action} is invariant under the BRST transformations
\begin{align}\label{eq:BRST_of_renorm_action}
 \mbox{ }\hspace{3.2cm}&sA_\mu&&\hspace*{-2ex}=D_\mu c\,, \mbox{ }\hspace{3cm} && s\,c &&\hspace*{-2ex}=\ri g{c}{c}\, ,\hspace{2.6cm}\mbox{ }\nonumber\\
 &s\,\bc&&\hspace*{-2ex}=b\,,                                                       && s\,b&&\hspace*{-2ex}=0\, ,  \nonumber\\
& s\,\bpsi_{\mu\nu}&&\hspace*{-2ex}=\bB_{\m\n}\,,                 && s\,\bB_{\m\n}&&\hspace*{-2ex}=0\,,\nonumber\\
& s\,B_{\m\n}&&\hspace*{-2ex}=\psi_{\m\n}\,,                    && s\,\psi_{\m\n}&&\hspace*{-2ex}=0\,,\nonumber\\
& s\, \bar{Q}_{\m\n\a\b}&&\hspace*{-2ex}=\bar{J}_{\m\n\a\b}, && s\, \bar{J}_{\m\n\a\b}&&\hspace*{-2ex}=0\,, \nonumber\\
& s\, Q_{\m\n\a\b}&&\hspace*{-2ex}=J_{\m\n\a\b}, && s\, J_{\m\n\a\b}&&\hspace*{-2ex}=0\,.
\end{align}
The auxiliary fields form BRST doublets reflecting their unphysical nature. 
Dimensions and ghost numbers of the fields involved are given in Table~\ref{tab:field_prop}.
\begin{table}[!ht]
\caption{Properties of fields and sources.}
\label{tab:field_prop}
\centering
\begin{tabular}{l c c c@{\hspace{9pt}} c@{\hspace{5pt}} c@{\hspace{5pt}} c@{\hspace{5pt}} c@{\hspace{5pt}} c@{\hspace{5pt}} c@{\hspace{5pt}} c@{\hspace{5pt}} c@{\hspace{5pt}} c@{\hspace{8pt}} c@{\hspace{11pt}} c@{\hspace{6pt}}}
\hline
\hline
\rule[12pt]{0pt}{0.1pt}
Field       & $A_\m$ & $c$ & $\bc$ & $B_{\m\n}$ & $\bB_{\m\n}$ & $\psi_{\m\n}$ & $\bpsi_{\m\n}$ & $J_{\a\b\m\n}$ & $\bar{J}_{\a\b\m\n}$ & $Q_{\a\b\m\n}$ & $\bar{Q}_{\a\b\m\n}$ & $\Omega^A_\m$ & $\Omega^c$ & $b$ \\[2pt]
\hline
$g_\sharp$  &    0   &  1  &   -1  &     0      &    0         &        1      &      -1        &   0            &       0              &  -1            &  -1 &   -1 &  -2  & 0 \\
Mass dim.   &    1   &  0  &   2   &     2      &    2         &        2      &       2        &   2            &       2              &  2             &  2 & 3 & 4 & 2\\
Statistics  &    b   &  f  &   f   &     b      &    b         &        f      &       f        &   b            &       b              &  f             &  f & f& b & b\\
\hline
\hline
\end{tabular}
\end{table}

The Slavnov-Taylor identity describing the BRST symmetry content of the model is given by
\begin{align}
\mathcal{B}(\Act)&=\intx\Bigg(\var{\Act}{\W^A_\m}\var{\Act}{A_\m}+\var{\Act}{\W^c}\var{\Act}{c}+b\var{\Act}{\bc}+\bB_{\m\n}\var{\Act}{\bpsi_{\m\n}}+\psi_{\m\n}\var{\Act}{B_{\m\n}}\nonumber\\*
&\phantom{=\intx\Bigg(} +\bar{J}_{\m\n\a\b}\var{\Act}{\bar{Q}_{\m\n\a\b}}+J_{\m\n\a\b}\var{\Act}{Q_{\m\n\a\b}}\Bigg)=0\,,
\end{align}
from which one derives the linearized Slavnov-Taylor operator
\begin{align}
\ST&=\intx\Bigg(\var{\Act}{\W^A_\m}\var{\ }{A_\m}+\var{\Act}{A_\m}\var{\ }{\W^A_\m}+\var{\Act}{\W^c}\var{\ }{c}+\var{\Act}{c}\var{\ }{\W^c}+b\var{\ }{\bc}\nonumber\\
&\phantom{=\intx\Bigg(} +\bB_{\m\n}\var{\ }{\bpsi_{\m\n}}+\psi_{\m\n}\var{\ }{B_{\m\n}}+\bar{J}_{\m\n\a\b}\var{\ }{\bar{Q}_{\m\n\a\b}}+J_{\m\n\a\b}\var{\ }{Q_{\m\n\a\b}}\Bigg)\,.
\end{align}
Furthermore we have the gauge fixing condition
\begin{align}
\var{\Act}{b}=\pa_\m A_\m=0\,,
\end{align}
the ghost equation
\begin{align}
\go(\Act)=\pa_\m\var{\Act}{\W^A_\m}+\var{\Act}{\bc}=0\,,
\end{align}
and the antighost equation
\begin{align}
\ag(\Act)=\intx\var{\Act}{c}=0\,.
\end{align}
Finally, we also have the symmetry $\U$:
\begin{align}
\U_{\a\b\m\n}(\Act)&=\intx\Bigg[B_{\a\b}\var{\Act}{B_{\m\n}}-\bB_{\m\n}\var{\Act}{\bB_{\a\b}}+J_{\a\b\r\s}\var{\Act}{J_{\m\n\r\s}}-\bar{J}_{\m\n\r\s}\var{\Act}{\bar{J}_{\a\b\r\s}}\nonumber\\
&\quad\qquad+\psi_{\a\b}\var{\Act}{\psi_{\m\n}}-\bpsi_{\m\n}\var{\Act}{\bpsi_{\a\b}}+Q_{\a\b\r\s}\var{\Act}{Q_{\m\n\r\s}}-\bar{Q}_{\m\n\r\s}\var{\Act}{\bar{Q}_{\a\b\r\s}}\Bigg]=0\,,
\end{align}
whose trace is connected to the reality of the action\footnote{In fact, the action is Hermitian since $\Q= 0$.}, and is hence denoted `reality charge' $\Q$~\cite{Vilar:2009}:
\begin{align}
\Q\equiv\d_{\a\m}\d_{\b\n}\U_{\a\b\m\n}\,.
\end{align}
Obviously, $\Q$ also generates a symmetry of the action \eqref{eq:renormalizable_action}. Having defined the operators $\ST$, $\ag$ and $\Q$ we may derive the following graded commutators:
\begin{align}\label{eq:algebra}
& \aco{\ag}{\ag}=0\,, & & \aco{\ST}{\ST}=0\,, & & \aco{\ag}{\ST}=0\,, \nonumber\\
& \co{\ag}{\Q}=0\,, & & \co{\Q}{\Q}=0\,, & & 
\co{\ST}{\Q}=0\,,
\end{align}
which means that these symmetry operators form a closed algebra. In fact, each of the relations in \eqnref{eq:algebra} gives rise to a constraint to possible counter terms in the effective action. However, first we will present the situation at one-loop level.

%
\section{Feynman Rules and Power Counting}
%
The simplest way to calculate the new gauge field propagator is to integrate over the auxiliary fields $B,\bB,\psi,\bpsi$ in the path integral after taking the physical values of $J,\bar{J},Q,\bar{Q}$ given in \eqnref{JQ-phys_new}:
\begin{align}
Z&=\int\mathcal{D}(\bpsi\psi\bB BA)\,\exp\bigg\{-\bigg(\Act_{\text{inv}}+\Act_{\text{gf}}+\Act_{\text{ext}}+\intx\bigg[\bpsi_{\m\n}\psi_{\m\n}-\bB_{\m\n}B_{\m\n}\nonumber\\
&\hspace{4.6cm} +\left(B_{\m\n}+\bar{B}_{\m\n}\right)\!\tfrac{\gamma^2}{2\wsq}\!\left(f_{\m\n}+\s\tfrac{\mth_{\m\n}}{2}\tilde{f}\right)\bigg]\bigg)\bigg\}\nonumber\\
&=\int\mathcal{D}(\bB BA)\,\exp\bigg\{-\bigg(\Act_{\text{inv}}+\Act_{\text{gf}}+\Act_{\text{ext}}+\intx\bigg[\tfrac{\gamma^4}{4\wsq}\!\left(f_{\m\n}+\s\tfrac{\mth_{\m\n}}{2}\tilde{f}\right)\tinv{\wsq}\!\left(f_{\m\n}+\s\tfrac{\mth_{\m\n}}{2}\tilde{f}\right)\nonumber\\
&\hspace{3.1cm} -\left(\bB_{\m\n}-\tfrac{\gamma^2}{2\wsq}\!\left(f_{\m\n}+\s\tfrac{\mth_{\m\n}}{2}\tilde{f}\right)\right)\left(B_{\m\n}-\tfrac{\gamma^2}{2\wsq}\!\left(f_{\m\n}+\s\tfrac{\mth_{\m\n}}{2}\tilde{f}\right)\right)\bigg]\bigg)\bigg\}\nonumber\\
&=\int\mathcal{D}A\exp\bigg\{-\bigg(\Act_{\text{inv}}+\Act_{\text{gf}}+\Act_{\text{ext}}+\intx\tfrac{\gamma^4}{4\wsq}\!\left(f_{\m\n}+\s\tfrac{\mth_{\m\n}}{2}\tilde{f}\right)\tinv{\wsq}\!\left(f_{\m\n}+\s\tfrac{\mth_{\m\n}}{2}\tilde{f}\right)\bigg)\bigg\}\,.
\end{align}
This leads to the action
\begin{align}
\Act_{\text{nl}}&=\intx\left(\inv{4}F_{\m\n}F_{\m\n}+\frac{\g^4}{4}\left[\inv{\wsq}f_{\m\n}\inv{\wsq}f_{\m\n}+\left(\s+\s^2\frac{\mth_{\m\n}\mth_{\m\n}}{4}\right)\inv{\wsq}\tilde{f}\inv{\wsq}\tilde{f}\right]+s\left(\bc\pa_\m A_\m\right)\right)\,,
\end{align}
which, using the abbreviation\footnote{Note, that from the special form of $\mth_{\m\n}$ in \eqnref{eq:def-eth} follows $\mth^2=4$, which is, however, not inserted at this point in order to keep the results more general.} $\mth^2=\mth_{\m\n}\mth_{\m\n}$, and with the definition of $\tilde{f}$, reduces to
\begin{align}
\Act_{\text{nl}}&=\intx\left(\inv{4}F_{\m\n}F_{\m\n}+\g^4\left[\pa_\m A_\n\inv{2\wsq^2} f_{\m\n}+\left(\s+\tfrac{\mth^2}{4}\s^2\right)(\tilde{\pa} A)\inv{\wsq^2}(\tilde{\pa} A)\right]+s\left(\bc\pa_\m A_\m\right)\right).
\end{align}

The gauge field propagator hence takes the form
\begin{align}
 G^{AA}_{\m\n}(k)&=\inv{k^2\left(1+\frac{\g^4}{(\k^2)^2}\right)}\left(\d_{\m\n}-\frac{k_\m k_\n}{k^2}- \frac{\st^4}{\left[\st^4+k^2\left(\k^2+\frac{\g^4}{\k^2}\right)\right]}\frac{\k_\m\k_\n}{\k^2}\right)\nonumber\\
&=\left[k^2+\frac{\g^4}{\k^2}\right]^{-1}\left[\d_{\m\n}-\frac{k_\m k_\n}{k^2}-\frac{\st^4}{\left(k^2+\left(\st^4+\g^4\right)\inv{\k^2}\right)}\frac{\k_\m\k_\n}{(\k^2)^2}\right]\,,
\label{eq:prop_aa}
\end{align}
where we have introduced the abbreviation
\begin{align}
 \st^4\equiv2\left(\s+\frac{\mth^2}{4}\s^2\right)\g^4\,,
\end{align}
and considered the case where $\mth_{\m\n}$ has the simple block diagonal form given in \eqref{eq:def-eth} so that $\k^2=k^2$. Two limits are of special interest: the IR limit $k^2\to0$ and the UV limit $k^2\to\infty$. A simple analysis reveals that
\begin{equation}
G^{AA}_{\m\n}(k)\approx\begin{cases}\frac{\k^2}{\g^4}\left[\d_{\m\n}-\frac{k_\m k_\n}{k^2}-\frac{\st^4}{\left(\st^4+\g^4\right)}\frac{\k_\m\k_\n}{\k^2}\right], & \text{for } \k^2\to0\,,\\[0.8em]
 \inv{k^2}\left(\d_{\m\n}-\frac{k_\m k_\n}{k^2}\right), & \text{for } k^2\to\infty\,.\\
\end{cases}
\label{eq:prop_aa_limits}
\end{equation}
From \eqnref{eq:prop_aa_limits} one can nicely see the appearance of a term of the same type as \eqref{eq:generic-IR-div} in the IR limit. This, by construction, admits the absorption of the problematic divergent terms appearing in the one loop results~\cite{Blaschke:2009d}. Another advantageous property of the gauge propagator is that the UV limit (from which due to {\uim} all divergences originate), admits to neglect the term proportional to $\g$ which reduces the number of terms in Feynman integrals considerably.

The ghost propagator takes the simple form
\begin{align}
G^{\bc c}(k)=-\frac{1}{k^2}\,,
\label{eq:prop_cbc}
\end{align}
which, as usual in a covariant gauge, is quadratically IR divergent. Alternatively, one could add a damping factor to the gauge fixing term and the ghost sector as has been done in Ref.~\cite{Blaschke:2009a}. However, such dampings appear in vertex expressions with an inverse power relative to the respective propagators and, thus, cancel each other. Moreover, these factors contribute to UV divergences at higher loop orders, and are omitted, hence.
Due to the missing coupling between the $A_\m$ and the remaining fields ($B,\bB,\psi,\bpsi$) no other propagator will contribute to physical results. However, for the sake of completeness, we give the respective expressions:
{\allowdisplaybreaks
\begin{subequations}\label{eq:otherprops}
\begin{align}
G^{BA}_{\m\n,\r}(k)&=\frac{\ri\g^2\left(k_\m\d_{\s\n}-k_\n\d_{\s\m}-\s\k_\s\mth_{\m\n}\right)}{2k^2\left(\k^2+\frac{\g^4}{\k^2}\right)}\left[\d_{\r\s}-\frac{\st^4}{\left[\st^4+k^2\left(\k^2+\frac{\g^4}{\k^2}\right)\right]}\frac{\k_\r\k_\s}{\k^2}\right]\nonumber\\
&=G^{\bB A}_{\m\n,\r}(k)\,,\label{eq:prop_BA}\\
G^{BB}_{\m\n,\r\s}(k)&=-\g^4\frac{\left(k_\m k_\r\d_{\n\s}+k_\n k_\s\d_{\m\r}-k_\m k_\s\d_{\n\r}-k_\n k_\r\d_{\m\s}\right)}{2k^2\k^2\left(\k^2+\frac{\g^4}{\k^2}\right)}\nonumber\\*
&\quad+\frac{\g^4}{4\k^2\left[k^2\left(\k^2+\frac{\g^4}{\k^2}\right)+\st^4\right]}\Bigg[\s\mth_{\m\n}\left(k_\r\k_\s-k_\s\k_\r\right)+\s\mth_{\r\s}\left(k_\m\k_\n-k_\n\k_\m\right)\nonumber\\
&\quad\qquad\;-\s^2\k^2\mth_{\m\n}\mth_{\r\s}-\st^4\frac{\left(k_\m\k_\n\k_\r k_\s+k_\r\k_\s\k_\m k_\n-k_\m\k_\n\k_\s k_\r+k_\s\k_\r\k_\m k_\n\right)}{k^2\k^2\left(\k^2+\frac{\g^4}{\k^2}\right)}\Bigg]\nonumber\\
&=G^{\bB\bB}_{\m\n,\r\s}(k),\label{eq:prop_BB}\\
G^{B\bB}_{\m\n,\r\s}(k)&=-\inv{2}\left(\d_{\m\r}\d_{\n\s}-\d_{\m\s}\d_{\n\r}\right)+G^{BB}_{\m\n,\r\s}(k)\,,\label{eq:prop_BbB}\\
G^{\bpsi\psi}_{\m\n\r\s}(k)&=-\inv{2}\left(\d_{\m\r}\d_{\n\s}-\d_{\m\s}\d_{\n\r}\right)\,.\label{eq:prop_Psi}
\end{align}
\end{subequations}
Notice, that all four propagators \eqref{eq:otherprops} tend towards a constant in the infrared as well as in the ultraviolet.

The model \eqref{eq:renormalizable_action} features three vertices which equal those of the `na\"ive' implementation of QED on {\nc} spaces (see for example Ref.~\cite{Hayakawa:1999b}):
\begin{subequations}
\begin{align}
\hspace{-3ex}\raisebox{-20pt}{\includegraphics[scale=0.8,trim=0 5pt 0 5pt,clip=true]{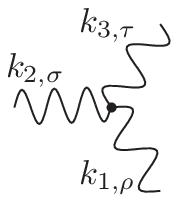}}&=\widetilde{V}^{3A}_{\rho\s\tau}(k_1, k_2, k_3)=2\ig(2\pi)^4\d^4(k_1+k_2+k_3)\sin\left(\tfrac{\sth}{2} k_1\k_2\right)\times\nonumber\\[-12pt]
&\phantom{=\widetilde{V}^{3A}_{\rho\s\tau}(k_1, k_2, k_3)\,}\quad\times\left[(k_3-k_2)_\rho \d_{\s\tau}+(k_1-k_3)_\s \d_{\rho\tau}+(k_2-k_1)_\tau \d_{\rho\s}\right],\label{eq:vert_3a}\\
\hspace{-3ex}\raisebox{-23pt}{\includegraphics[scale=0.8,trim=0 2pt 0 5pt,clip=true]{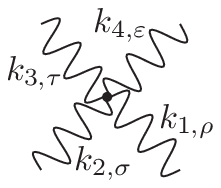}}\hspace{-9pt}&=\widetilde{V}^{4A}_{\rho\s\tau\e}(k_1, k_2, k_3, k_4)=-4g^2(2\pi)^4\d^4(k_1+k_2+k_3+k_4)\times\nonumber\\*[-16pt]
&\phantom{=\widetilde{V}^{4A}_{\rho\s\tau\e}(k_1, k_2, k_3, k_4)=}\;\times\Big[(\d_{\rho\tau}\d_{\s\e}-\d_{\rho\e}\d_{\s\tau})\sin\left(\tfrac{\sth}{2}k_1\k_2\right)\sin\left(\tfrac{\sth}{2}k_3\k_4\right)\nonumber\\*
&\phantom{=\widetilde{V}^{4A}_{\rho\s\tau\e}(k_1, k_2, k_3, k_4)=}\;\phantom{\Big[}+(\d_{\rho\s}\d_{\tau\e}-\d_{\rho\e}\d_{\s\tau})\sin\left(\tfrac{\sth}{2}k_1\k_3\right)\sin\left(\tfrac{\sth}{2}k_2\k_4\right)\nonumber\\*
&\phantom{=\widetilde{V}^{4A}_{\rho\s\tau\e}(k_1, k_2, k_3, k_4)=}\;\phantom{\Big[}+(\d_{\rho\s}\d_{\tau\e}-\d_{\rho\tau}\d_{\s\e})\sin\left(\tfrac{\sth}{2}k_2\k_3\right)\sin\left(\tfrac{\sth}{2}k_1\k_4\right)\Big],\label{eq:vert_4a}\\
\hspace{-3ex}\raisebox{-23pt}{\includegraphics[scale=0.8,trim=0 2pt 0 5pt,clip=true]{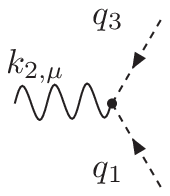}}\hspace{0.1ex}&=\widetilde{V}^{\bc Ac}_\mu(q_1, k_2, q_3)=-2\ig(2\pi)^4\d^4(q_1+k_2+q_3)q_{3\mu}\sin\left(\tfrac{\sth}{2}q_1\q_3\right)\label{eq:vert_cAbc}\,.
\end{align}
\end{subequations}
}

Concerning the superficial degree of UV divergence $d_\gamma$ one can set up the following relations for the number of loops $L$, external lines $E_\phi$, internal lines $I_\phi$ and vertices $V_{\phi}$ for fields $\phi\mathrel{\in}\{A,c,\bc\}$:
\begin{align}
 L&=I_A+I_{c\bc}-\left(V_{\bc Ac}+V_{3A}+V_{4A}-1\right)\,,\nonumber\\
E_{c\bc}+2I_{c\bc}&=2V_{\bc Ac}\,,\nonumber\\
E_A+2I_A&=3V_{3A}+4V_{4A}+V_{\bc Ac}\,,\\
\intertext{and counting the UV powers of respective Feynman rules yields}
d_\gamma&=4L-2I_A-2I_{c\bc}+V_{3A}+V_{\bc Ac}\,.
\end{align}
This system of equations can be resolved by eliminating the $I_\phi$ and $V_{\phi}$, finally leading to
\begin{align}\label{eq:powercounting}
 d_\gamma=4-E_A-E_{c\bc}\,,
\end{align}
which, again, shows remarkable agreement with the respective relations for the `na\"{\i}ve' implementation of {\nc} $U_\star(1)$ gauge theory.

%
\section{One-loop Calculations}\label{sec:one-loop}
%
\subsection{Vacuum Polarization}\label{sec:vac-pol}
\begin{figure}[!ht]
 \centering
 \includegraphics[scale=0.8]{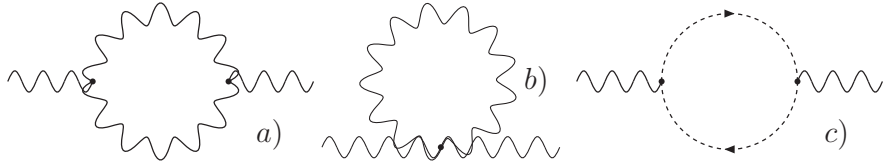}
 \caption{One loop corrections to the gauge boson propagator.} 
 \label{fig:1loop_vacpol_all}
\end{figure}

%
The Feynman rules \eqref{eq:prop_aa},\eqref{eq:prop_cbc},\eqref{eq:prop_BB}--\eqref{eq:prop_BbB}, and \eqref{eq:vert_3a}--\eqref{eq:vert_cAbc} give rise to the three graphs in \figref{fig:1loop_vacpol_all} contributing to the vacuum polarization $\Pi_{\m\n}(p)$.

In the light of renormalization one is interested mainly in the divergence structure of these expressions in the limit of small external momenta $p^2\to0$. Therefore, we apply an expansion of the following form to the integrands $\mathcal{I}(k)$,
\begin{align}
 \label{eq:1-l_expansion}
\Pi_{\mu\nu}(p)=& \intk\, \mathcal{I}_{\mu\nu}(p,k) \sin^2 \left( \frac{\sth}{2} k\p \right)\nonumber\\
\approx& \intk\sin^2 \left( \frac{\sth}{2} k\p \right) \bigg[\mathcal{I}_{\mu\nu}(0,k) +p_\rho \left[\partial_{p_\rho}\mathcal{I}_{\mu\nu}(p,k)\right]_{p\to0} \nonumber\\
&\hspace{2.8cm}+\frac{p_\rho p_\s}{2} \left[\partial_{p_\rho}\partial_{p_\s}\mathcal{I}_{\mu\nu}(p,k)\right]_{p\to0} +\mathcal{O} \left(p^3\right)\bigg] .
\end{align}
The phase factors are not expanded in order not to lose the damping effect of the highly oscillating functions at high momenta $k$. Notice, that since all potential divergences (UV as well as IR coming from {\uim}) arise from the region of \emph{large} internal momentum $k$, one may use the UV approximation for the gauge field propagator \eqnref{eq:prop_aa_limits} for all one-loop calculations. Conducting the expansions and integrations\footnote{For mathematical details of the calculations please see the recent publications \cite{Blaschke:2009a, Blaschke:2008b}.} for the graphs depicted in \figref{fig:1loop_vacpol_all} one finally arrives at
\begin{align}
\Pi_{\mu\nu}(p)=&\frac{2 g^2}{\pi^2\sth^2}\frac{\p_\m \p_\n}{\left(\p^2\right)^2}+\frac{13 g^2}{3 (4\pi)^2}\left(p^2\delta_{\m\n}-p_\m p_\n\right)\ln\left(\L\right)+\text{finite terms}\,,
\label{eq:res_vac_pol_divergence}
\end{align}
where $\L$ denotes an ultraviolet cutoff, and `finite terms' collects contributions being finite in the limits $\L\to\infty$ and $\p^2\to0$, respectively. As expected this result exhibits a quadratic IR divergence showing the tensor structure of \eqnref{eq:generic-IR-div}.

%
\subsection{Vertex Corrections}
%
\begin{figure}[!ht]
 \centering
 \includegraphics[scale=0.8]{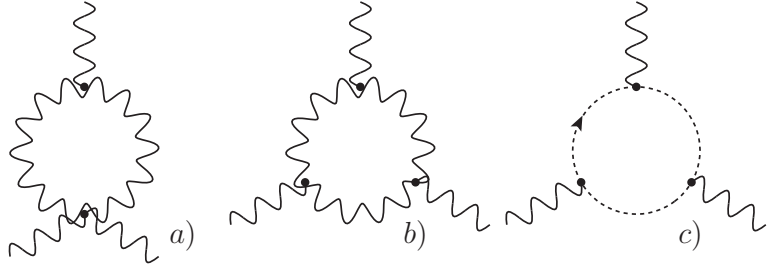}
 \caption{One loop corrections to the 3A-vertex.} 
 \label{fig:1loop_3A_all}
\end{figure}
In principal, the calculation of the vertex corrections proceeds along the lines of the previous section. However, one has to guarantee that the limits $p_i\to 0$, $i=1,2,3$ commute for the integrand and its derivatives. A typical divergent expression, as it appears in the calculation of the graphs depicted in \figref{fig:1loop_3A_all}, reads
\begin{align}
 \mathcal{I}_{\m\n\r}(k,p_1,p_2,p_3)=\!\intk\, \Xi_{\s\t\xi}(p_{1,\m},p_{2,\n},p_{3,\r})\frac{k_\s k_\t k_\xi}{(\sth k^2)^3}\sin\!\left(\sth\tfrac{p_1(\p_2-\k)}{2}\right) \sin\!\left(\sth\tfrac{p_2\k}{2}\right) \sin\!\left(\sth\tfrac{p_3\k}{2}\right)\!,
\end{align}
where $\Xi$ denotes a regular smooth function. In order to evaluate this integral it is convenient to utilize the identity
\begin{align}
&\sin\!\left(\sth \frac{p_1(\p_2-\k)}{2}\right)\sin\!\left(\sth \frac{p_2\k}{2}\right)\sin\!\left(\sth\frac{p_3\k}{2}\right)\nonumber\\
& = \inv{4}\cos\!\left(\sth \frac{p_1 \p_2}{2}\right)\left(\sin\!\left(\sth p_1\k\right) +\sin\!\left(\sth p_2\k\right) +\sin\!\left(\sth p_3\k\right)\right)\nonumber\\
&\quad-\inv{4}\sin\!\left(\sth\frac{p_1\p_2}{2}\right)\left(1+\cos\!\left(\sth p_1\k\right)-\cos\!\left(\sth p_2\k\right) - \cos\!\left(\sth p_3 \k\right)\right)\,.
\end{align}
The sum of the graphs depicted in \figref{fig:1loop_3A_all} results in a linear IR divergence of the form
\begin{align}
\label{eq:3-A correction_IR}
\Gamma^{3A,\text{IR}}_{\m\n\r}(p_1,p_2,p_3)&=-\frac{2\ri g^3}{\pi^2}\cos\left(\sth \frac{p_1\p_2}{2}\right)\sum\limits_{i=1,2,3}\frac{\p_{i,\m}\p_{i,\n}\p_{i,\r}}{\sth(\p_i^2)^2}\,,
\end{align}
as well as a logarithmic UV divergence
\begin{align}
\label{eq:3-A correction_UV}
\Gamma^{3A,\text{UV}}_{\m\n\r}(p_1,p_2,p_3)&=\frac{17}{3}\ig^3\pi^2\ln(\L)\sin\left(\sth \frac{p_1\p_2}{2}\right)\Big[(p_1-p_2)_\r\d_{\m\n}+(p_2-p_3)_\m\d_{\n\r}\nonumber\\
&\hspace{4.9cm}+(p_3-p_1)_\n\d_{\m\r}\Big]\nonumber\\
&=-\frac{17\,g^2}{6(4\pi)^2}\ln (\Lambda) \widetilde{V}^{3A,\text{tree}}_{\m\n\r}(p_1,p_2,p_3)\,,
\end{align}
where due to momentum conservation $p_3=-p_1-p_2$. We note that our results \eqref{eq:3-A correction_IR} and \eqref{eq:3-A correction_UV} resemble the ones given in the literature \cite{Matusis:2000jf, Armoni:2000xr,Ruiz:2000}.

\begin{figure}[!ht]
 \centering
 \includegraphics[scale=0.8]{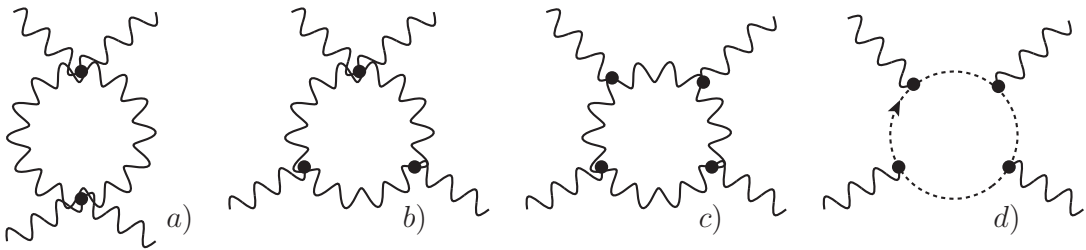}
 \caption{One-loop corrections to the 4A-vertex.} 
 \label{fig:1loop_4A_all}
\end{figure}
The Feynman graphs contributing to the one-loop 4-A vertex corrections are depicted in \figref{fig:1loop_4A_all}. They are all IR finite. 
Gauge invariance and the renormalization of the coupling constant from the 3-A vertex, as discussed in Section \ref{sec:betafun}, lead to the following logarithmic UV divergence in the 4-A vertex correction (cf.~\cite{Martin:2000bk}):
\begin{align}
\label{eq:4-A correction_UV}
\Gamma^{4A,\text{UV}}_{\m\n\r\s}(p_1,p_2,p_3,p_4)& =
- \frac{5}{8\pi^2} \ln(\L) g^2\, \widetilde{V}^{4A,\text{tree}}_{\m\n\r\s}(p_1,p_2,p_3,p_4), 
\end{align}
where due to momentum conservation $p_4=-p_1-p_2-p_3$.

%
\subsection{Renormalization}
\label{sec:renorm}

For a renormalizable theory, the form of the action is invariant under quantum corrections (provided the subtraction scheme respects all the symmetries) and the parameters are fixed by renormalization conditions on the vertex functions.
Recall that the tree-level gauge field propagator has the form
\begin{align}\label{eq:propAA-general}
G^{AA}_{\m\n}(k)&=\inv{k^2\mathcal{D}}\left(\d_{\m\n}-\left(1-\a\mathcal{D}\right)\frac{k_\m k_\n}{k^2}-\mathcal{F}\frac{\k_\m\k_\n}{\k^2}\right)\,,
\end{align}
where we have introduced the abbreviations
\begin{align}
\mathcal{D}(k)&\equiv\left(1+\frac{\g^4}{(\k^2)^2}\right)\,, \nonumber\\
\mathcal{F}(k)&\equiv\inv{\k^2}\frac{\st^4}{\left(k^2+\left(\st^4+\g^4\right)\inv{\k^2}\right)}\,.
\end{align}
So far, we have worked in Landau gauge with $\a=0$ in order to simplify loop-calculations (cf. \eqnref{eq:prop_aa}). However, in the following a more general gauge fixing with arbitrary gauge parameter $\a$ will be advantageous\footnote{Note, that the quadratic IR divergence is independent of the gauge fixing~\cite{Blaschke:2005b,Ruiz:2000,Hayakawa:1999b}.}, as we explicitly need the inverse of the propagator \eqref{eq:propAA-general}, which for obvious reasons diverges for $\a\to0$ due to the elimination of the Lagrange multiplier field $b$:
\begin{align}
 \G^{AA,\text{tree}}_{\m\n}(k)&=\left(G_{AA}^{-1}\right)_{\m\n}(k)=k^2\mathcal{D}\left(\d_{\m\n}+\left(\inv{\a\mathcal{D}}-1\right)\frac{k_\m k_\n}{k^2}+\frac{\st^4}{k^2\k^2\mathcal{D}}\frac{\k_\m\k_\n}{\k^2}\right)\,.
\label{eq:AA-tree-vertex}
\end{align}
Of course, one could do these computations also in the Landau gauge, but one would have to take $b$ and its (mixed) propagator into account. Instead, in order to keep things simple, we will take the limit $\a\to0$ only in the final step. 

In Section~\ref{sec:vac-pol} we have computed the one-loop corrections to the tree-level two-point vertex function \eqref{eq:AA-tree-vertex}. It is given by \eqnref{eq:res_vac_pol_divergence}, i.e. 
\begin{align}
 \G^{AA,\text{corr.}}_{\m\n}(k)&=\varPi_1\frac{\k_\m\k_\n}{(\k^2)^2}+\varPi_2\left(k^2\d_{\m\n}-k_\m k_\n\right)\,,\nonumber\\
\varPi_1&=\frac{2g^2}{\pi^2\sth^2} \,,
\qquad \varPi_2=\frac{13g^2}{3(4\pi)^2}\ln\L\,,
\end{align}
where $\L$ was an ultraviolet cutoff. 
Hence, with these definitions for $\varPi_i$ we find
\begin{align}
 \G^{AA,\text{ren}}_{\m\n}(k)&=\G^{AA,\text{tree}}_{\m\n}(k)-\G^{AA,\text{corr.}}_{\m\n}(k)\nonumber\\
&=k^2(\mathcal{D}-\varPi_2)\left(\d_{\m\n}+\left(\inv{\a(\mathcal{D}-\varPi_2)}-1\right)\frac{k_\m k_\n}{k^2}+\frac{\st^4-\varPi_1}{k^2\k^2(\mathcal{D}-\varPi_2)}\frac{\k_\m\k_\n}{\k^2}\right)\,.
\end{align}
Introducing the wave-function renormalization $Z_A$ and the renormalized parameters $\g_r$ and $\st_r$ according to
\begin{align}
\label{ZA}
 Z_A&=\inv{\sqrt{1-\varPi_2}}\,,\nonumber\\
\g_r^4&={\g^4}{Z_A^2}\,,\nonumber\\
\st_r^4&=\left({\st^4-\varPi_1}\right){Z_A^2}\,,
\end{align}
the one-loop two-point vertex function is cast into the same form as its tree-level counter part, i.e.
\begin{align}
 \G^{AA,\text{ren}}_{\m\n}(k)&=\frac{k^2\mathcal{D}_r}{Z_A^2}\left(\d_{\m\n}+\left(\frac{Z_A^2}{\a\mathcal{D}_r}-1\right)\frac{k_\m k_\n}{k^2}+\frac{\st_r^4}{k^2\k^2\mathcal{D}_r}\frac{\k_\m\k_\n}{\k^2}\right)\,,\nonumber\\
\mathcal{D}_r(k)&\equiv\left(1+\frac{\g_r^4}{(\k^2)^2}\right)\,.
\end{align}
Finally, we may also write $\st_r$ in terms of the renormalized $\s_r$:
\begin{align}
\st_r^4&=2\left(\s_r+\frac{\mth^2}{4}\s_r^2\right){\g^4}{Z_A^2}\,,\nonumber\\
\s_r&= -\frac{2}{\th^2}\pm2\sqrt{\left(1+\frac{\mth^2}{2} \s\right)^2-\frac{g^2\mth ^2}{\pi^2\g^4 \sth^2}}\ .
\end{align}
For the sake of completeness, we note that the renormalized propagator in Landau gauge ($\a\to0$) becomes
\begin{align}
G^{AA,\text{ren}}_{\m\n}(k)&=\frac{Z_A^2}{k^2\mathcal{D}_r}\left(\d_{\m\n}-\frac{k_\m k_\n}{k^2}-\mathcal{F}_r\frac{\k_\m\k_\n}{\k^2}\right)\,,\nonumber\\
\mathcal{F}_r&\equiv\inv{\k^2}\frac{\st_r^4}{\left(k^2+\left(\st_r^4+\g_r^4\right)\inv{\k^2}\right)}\,.
\end{align}

In the following, we will provide renormalization conditions for the two-point vertex function for the gauge boson
\begin{align}
\nonumber
\Gamma^{AA}_{\mu\rho} 
& = \Gamma^{AA,T} (\d_{\m\r} - \frac{k_\m k_\r}{k^2}) +(\Gamma^{AA,NC})\, \frac{\tilde k_\m \tilde k_\r}{\tilde k^2}  + (\Gamma^{AA,L})\, \frac{k_\m k_\r}{k^2}\,,
\end{align}
where the vertex function has been split into a transversal and longitudinal part. We have used the identifications
\begin{align}
\Gamma^{AA,T} & = k^2 \mathcal D\,,\qquad
\Gamma^{AA,NC} =  \frac{\st^4}{\k^2}\,,\qquad (\Gamma^{AA,L}) = \frac{k^2}{\alpha} \,,
\end{align}
which finally allow to formulate the following renormalization conditions:
\begin{subequations}
\begin{align}
\label{cond-1}
\frac{(\k^2)^2}{k^2} \Gamma^{AA,T} \Big|_{k^2=0} & = \gamma^4\,,\\
\label{cond-2}
\frac 1{2 k^2} \frac{\partial (k^2 \Gamma^{AA,T})}{\partial k^2} \Big|_{k^2=0} & = 1\,,\\
\label{cond-3}
\k^2 \Gamma^{AA,NC}\Big|_{k^2=0} & = \bar \sigma^4\,,\\
\label{cond-4}
\Gamma^{AA,L}\Big|_{k^2=0} & = 0\,,\\
\label{cond-5}
\frac{\partial \Gamma^{AA,L}}{\partial k^2} \Big|_{k^2=0} & = \frac 1 \alpha\,.
\end{align}
\end{subequations}

Regarding the renormalization arising from the three-point functions\footnote{Remember that the four-point functions are IR finite.} the IR divergent result \eqref{eq:3-A correction_IR} corresponds to the counter term
\begin{align}
\label{3A-correction}
\Act^{3A,corr}=\intx g^3\aco{A_\m}{A_\n} \frac{\tilde{\pa}_\m\tilde{\pa}_\n\tilde{\pa}_\r}{\sth\, \wsq^2}A_\r\,.
\end{align}
Such a term can readily be introduced into the soft-breaking part of the action $\Act_\text{soft}$ in \eqnref{eq:renormalizable_action}. But in order to do so, we have to restore BRST invariance in the UV regime. Again, this can be achieved by introducing sources $Q'$ and $J'$, which form a BRST doublet,
\begin{align}
sQ' = J'\,, \qquad sJ'=0\,.
\end{align}
Consequently, we insert the following terms into $\Act_\text{soft}$ in \eqnref{eq:renormalizable_action}:
\begin{align}
\label{counter-3A}
\intx \left[J'\aco{A_\m}{A_\n} \frac{\tilde{\pa}_\m\tilde{\pa}_\n\tilde{\pa}_\r}{\wsq^2}A_\r -
Q's\left(\aco{A_\m}{A_\n} \frac{\tilde{\pa}_\m\tilde{\pa}_\n\tilde{\pa}_\r}{\wsq^2}A_\r\right)\right]\,.
\end{align}
This term is BRST invariant, \emph{per se}. In the IR, the sources take on their physical values
\begin{align}
J' = g\g'^2,\,\, 
Q'= 0\,,
\end{align}
(cf. \eqnref{JQ-phys_new} and Refs.~\cite{Zwanziger:1993,Dudal:2008,Vilar:2009}), and the counter term in \eqnref{3A-correction} leads to a renormalized $\g'$, which is another parameter of dimension 1. Adding the term \eqref{counter-3A} to the action does not alter the divergence structure of the theory, due to the damping behavior of the propagators in the IR, and the fact that the additional contribution to the 3A-vertex is subleading in the UV, i.e. it does not modify the UV power counting \eqref{eq:powercounting}.  

In contrast to the IR singular terms, the UV divergences of \eqref{eq:3-A correction_UV} and \eqref{eq:4-A correction_UV} can be absorbed into the coupling constant, which is discussed subsequently in \secref{sec:betafun}.

\subsection{The \texorpdfstring{$\beta$}{beta}-Function}
%
\label{sec:betafun}
As usual, the $\beta$-function is given by the logarithmic derivative of the coupling $g$ with respect to the cut-off for fixed $g_r$:
\begin{eqnarray}
\b(g,\L) & = & \L \frac{ \pa g }{ \pa \L}\Big|_{g_r\, \text{fixed}}\,,\label{eq:betafun_def}
\nonumber\\
\b(g) & = & \lim_{\L\to\infty} \beta(g,\L)\,.
\end{eqnarray}
The renormalized coupling is obtained from the relation
\begin{align}
\label{gg}
g_r = g Z_g Z_A^3\,,
\end{align}
where $Z_g$ denotes the multiplicative correction to the three-photon vertex \eqref{eq:3-A correction_UV}, and $Z_A$ is the wave function renormalization given in \eqref{ZA}. Explicitly, we have
\begin{subequations}
\label{eq:betafun_explicit_corr}
\begin{align}
Z_g &= 1 + \frac {17\,g^2}{96\,\pi^2 }\ln \Lambda\,, \label{eq:betafun_zg_explicit}\\
Z_A &= \left( 1 - \frac{13 g^2}{3 (4\pi)^2} \ln \Lambda\right)^{-1/2}\,.\label{eq:betafun_za_explicit}
\end{align}
\end{subequations}
Note that the appearance of the wave function renormalization $Z_A$ in \eqnref{gg} is reasoned by the vertex correction being computed with the unrenormalized fields $A_\m$, which need to be replaced by their renormalized counter parts $A^r_\mu = Z^{-1}_A A_\mu$. Since this topic is treated rather novercally in the literature, let us briefly sketch how the $\beta$-function is obtained in practise.

Bearing in mind that, according to \eqnref{eq:betafun_def}, the renormalized coupling is being held constant, we differentiate \eqref{gg} (after expansion for small $g$) with respect to $\Lambda$ and multiply the result by $\Lambda$. With the general definition $g_r = g(1+g^2 v \ln \Lambda) + \mathcal O(g^5)$, this yields
\begin{align}
\label{eq:betafun_tosolve}
0 = \beta(g,\Lambda) + 3 \, v   g^2 \beta(g,\Lambda) \ln\Lambda + vg^3\, .
\end{align}
The solution of \eqref{eq:betafun_tosolve} is then obtained by the ansatz 
\[
\beta(g, \Lambda) = - g^3 v + \mathcal O(g^5)\,,
\]
and we finally obtain
\begin{align}
\beta(g)  = \lim_{\Lambda\to\infty} \beta(g,\Lambda) = - g^3 v\,.
\end{align}
Inserting the explicit definitions \eqref{eq:betafun_explicit_corr}, the $\beta$-function reads
\begin{align}
\label{beta}
\beta(g) =- \frac{7}{12}\frac{g^3}{\pi^2}\,  <\, 0\,.
\end{align}
The negative sign of the $\beta$-function indicates asymptotic freedom and the absence of a Landau ghost. This qualitatively compares to the results in \cite{Martin:1999aq, Armoni:2000xr,Minwalla:1999,Ruiz:2000}, where also a negative $\beta$-function\footnote{but with different numerical factors} is obtained for the case of $\th$-deformed QED.

\section{IR Damping in Higher Loop Order}\label{sec:higher-loops}
%
In the light of renormalizability at higher loop orders, it is important to investigate the IR behavior of
such integrands with insertions of the one-loop corrections which were discussed in the previous section. The aim is to identify possible poles\footnote{This endeavor is motivated by the fact, that the one loop graphs of \secref{sec:one-loop} give rise to IR divergences. Upon insertion of these functions into higher-loop graphs the former external momenta are being integrated over and, hence, also touch the value $\p^2=0$.} at $\p^2 = 0$. Hence, we consider a chain of $n$ non-planar insertions, which may be part of a higher loop graph. 
\begin{figure}[!ht]
  \centering
 \includegraphics[scale=0.64]{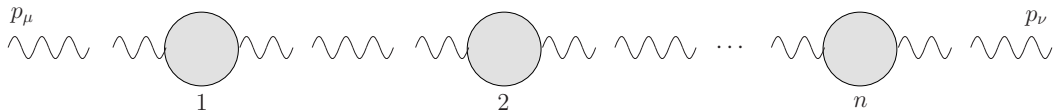}
 \caption{A chain of $n$ non-planar insertions, concatenated by gauge field propagators.}
 \label{fig:nloop_AA}
\end{figure}

{\noindent}Considering, for example, the IR behavior of the chain depicted in \figref{fig:nloop_AA} we find by employing Eqns. \eqref{eq:prop_aa_limits} and \eqref{eq:res_vac_pol_divergence}:
\begin{align}\label{higher-loop-chain}
\left(G^{AA}(p)\Pi^{\text{np}}(p)\right)^n_{\m\s}G^{AA}_{\s\n}(p)\Big|_{\text{IR}}\approx \frac{g^{2n}}{\left[\sth^2\left(\st^4+\g^4\right)\right]^{n+1}}\p_\m\p_\n\,,
\end{align}
i.e. the expression is well-behaved in the limit of small $\p$. Therefore, no IR problems are expected in higher loop graphs with insertions of the above type. In fact, \eqref{higher-loop-chain} represents a nice illustration of the damping mechanism implemented by the soft breaking term in the action \eqref{eq:renormalizable_action} and parametrized by $\g$. In the limit $\g\to0$ (implying $\st\to0$) the divergent behavior of the `na\"ive' model without soft breaking is recovered.

%
\section{Discussion and Outlook}\label{sec:conclusion}
%
After suffering a setback in our enterprise to construct a renormalizable {\nc} gauge field model~\cite{Blaschke:2009c} we have thoroughly analyzed the obstacles raised by the introduction of a parameter of non-commutativity with negative mass dimension~\cite{Blaschke:2009d}. 
Benefiting from the insight gained from this analysis we have worked out a {\nc} $U_\star(1)$ gauge model which provides the necessary terms to absorb all divergences appearing at one-loop level. The main concepts of the construction are:
\begin{itemize}
\item [-] The IR damping of the $1/p^2$ model is implemented by a soft breaking term. This allows us to use ordinary derivatives instead of inverse covariant derivatives which need to be localized by auxiliary fields and ghosts in order to be interpreted in a reasonable way. Thereby, the gauge field $A$ is completely decoupled (in terms of interactions) from the new auxiliary fields, and respective quantum corrections reduce (diagrammatically) to the ones known from the na\"ive non-commutative YM and QED theories~\cite{Hayakawa:1999b, Matusis:2000jf}. 
\item [-] The tree level action features terms $\propto (\tilde\partial A)^n$, with powers $n=2,3$ being designed to absorb all types of divergences appearing at one-loop level. From transversality with respect to external momenta, dimensional considerations, and power counting, one can derive that these, in fact, are the only IR divergent tensor structures appearing in the corrections.
\end{itemize}
Explicit one-loop computations showed the expected results. The vacuum polarization of the gauge boson behaves for small external momenta as
\begin{align}
\Pi^{\text{IR}}_{\m\n}(p)\propto \frac{\p_\m\p_\n}{\sth^2(\p^2)^2}\,,
\end{align}
a divergence which has been shown to be absorbable into the parameter of the $\inv{\wsq^2}(\tilde{\partial}A)^2$ term in the action. The correction of the $3A$ vertex reads
\begin{align}
\Gamma^{3A,\text{IR}}_{\m\n\r}(p_1,p_2,p_3)\propto\cos\left(\sth p_1\p_2\right)\sum\limits_{i=1,2,3}\frac{\p_{i,\m}\p_{i,\n}\p_{i,\r}}{\sth(\p_i^2)^2}\,,
\end{align}
corresponding to the renormalization of the (massive) parameter of the term $\inv{\sth\wsq^2}(\tilde{\partial}A)^3$. 
These one-loop results lead to a negative $\beta$-function (\ref{beta}), indicating asymptotic freedom and the absence of a Landau ghost. This remarkable fact is due to the `non-Abelian' nature of the non-commutative $U_\star(1)$ theory.

Although, at the moment, we have not been able to construct a rigorous proof, there are strong indications that the properties mentioned above indeed lead to renormalizability of this model. We may conjecture that this can be confirmed in the framework of Multiscale Analysis~(for a review see Ref.~\cite{Rivasseau:2007a}). This, however, will require extensive work to establish the foundations~\cite{Rivasseau:1991ub} for the application of this scheme to gauge theories.

\subsection*{Acknowledgements}
The authors are indebted to H.~Grosse, F.~Heindl, E.~Kronberger and H.~Steinacker for valuable discussions. The work of D.~N.~Blaschke, A.~Rofner, R.~I.~P.~Sedmik and M.~Wohlgenannt was supported by the `Fonds zur F\"orderung der Wissenschaftlichen Forschung' (FWF) under contracts P20507-N16 and P21610-N16.



\end{document}